\documentclass[epj,referee]{svjour}
\usepackage{graphicx}
\newcommand{\be}{\begin{equation}}
\newcommand{\ee}{\end{equation}}
\newcommand{\bea}{\begin{eqnarray}}
\newcommand{\eea}{\end{eqnarray}}

\newcommand{\fra}[2]{\hbox{${#1\over #2}$}}
\begin{document}
\title{Asymptotic and effective coarsening exponents in surface growth models}
\author{Paolo Politi$^{(1)}$ \and Alessandro Torcini$^{(1,2)}$
}                     
\mail{Paolo.Politi@isc.cnr.it}
\institute{(1) Istituto dei Sistemi Complessi, 
Consiglio Nazionale delle Ricerche,
Via Madonna del Piano 10, 50019 Sesto Fiorentino, Italy \\
(2) Istituto Nazionale di Fisica Nucleare, Sezione di Firenze, via Sansone 1, 50019 Sesto Fiorentino, Italy}
\date{Received: date / Revised version: date}
%
\abstract{
We consider a class of unstable surface growth models,
$\partial_t z = -\partial_x {\cal J}$,
developing a mound structure of size $\lambda$ and displaying a
perpetual coarsening process, i.e. an endless increase in time of $\lambda$.
The coarsening exponents $n$,
defined by the growth law of the mound size $\lambda$ with time,
$\lambda \sim t^n$, were previously found by numerical integration of the
growth equations [A. Torcini and P. Politi, Eur. Phys. J. B 25, 519 (2002)]. 
Recent analytical work 
now allows to interpret such findings as finite time effective
exponents. The asymptotic exponents are shown to appear at so large
time that cannot be reached by direct integration of
the growth equations. The reason for the appearance of effective exponents
is clearly identified.
\PACS{
{68}{Surfaces and interfaces} \and
{81.10.Aj}{Theory and models of crystal growth} \and
{02.30.Jr}{Partial differential equations}
     } 
} 
\maketitle
\section{Introduction}
\label{intro}

In this manuscript we are interested in studying a class 
of one dimensional growth equations, having the conserved form $\partial_t z
= F -\partial_x {\cal J}$: the dynamics of the local height $z(x,t)$ of the
surface is determined, apart from a trivial constant term $F$ describing
the deposition flux, by the processes occuring at the surface,
which are all included in the surface current 
${\cal J}$~\cite{review,libro_JV}.
Possible noise sources, shot noise first, will be neglected.

The current ${\cal J}$ may have a plethora of different forms, depending
on the details of the atomistic processes~\cite{proceeding}.
For Molecular Beam Epitaxy~\cite{Evans,libro_MK}, 
a widely used technique for growing metal and semiconductor
thin films with nanoscale control, one of the most studied equations
has the form~\cite{review,libro_MK}
\be
\partial_t z = -\partial_x (u_{xx} + j(u) )
\label{eq_z}
\ee
where the constant term $F$ has been included in the left hand side
by redefining $z=z-Ft$, and $u=z_x=\partial_x z$ is the slope 
of the surface.
It is worth noting that taking the spatial derivative of both sides,
we get $u_t = -\partial_{xx} (u_{xx} + j(u))$, i.e. a generalized Cahn-Hilliard
equation~\cite{PRE}. Without the double derivative $(-\partial_{xx})$
we get the corresponding non-conserved models, called
generalized real Ginzburg-Landau (or Allen-Cahn) equation~\cite{PRE}.
In summary, we are considering the two classes of equations
\bea
u_t = -\partial_{xx} (u_{xx} +j(u)) && \mbox{conserved}\label{gCH} \\
u_t = u_{xx} +j(u) && \mbox{non-conserved}\label{gGL} 
\eea

The linear stability analysis of the flat
interface, $z =z_0 + \epsilon \exp(\omega t + iqx)$, 
can be easily applied to Eqs.~(\ref{gCH},\ref{gGL}) giving
\bea
\omega = j'(0)q^2 - q^4 && \mbox{conserved}\\
\omega = j'(0) - q^2 && \mbox{non-conserved},
\eea
so that
an instability appears (i.e. $\omega(q)>0$ for some $q$) if $j'(u=0)>0$.
Without loss of generality, we may assume $j'(0)=1$.
The steady states are determined by the equation $u_{xx}=-j(u)$
both in the conserved (\ref{gCH}) and non-conserved (\ref{gGL}) case.
The solutions $u(x)$ correspond to the trajectories of a fictitious 
particle moving in the symmetric potential
$V(u)=\int du\, j(u) = \fra{1}{2}u^2 +$ {\em higher order terms}.
The oscillations in the potential well correspond
to periodic steady states of the variable $u$, with wavelength 
$\lambda$ and amplitude $A$.

The study of the stationary periodic solutions is of great
importance for the dynamics as well. In Ref.~\cite{PRL},
for Eqs.~(\ref{gCH}-\ref{gGL}) and other classes of models,
it has been shown that the surface undergoes a coarsening process
if and only if $d\lambda/dA >0$.
In simple words, the wavelength of the mound structure (emerging
from the linear instability) increases in time if the wavelength
$\lambda(A)$ of the periodic steady state increases with the
amplitude $A$. Even more importantly, the knowledge of the
stationary periodic solutions allows to determine~\cite{PRE}
the coarsening law $\lambda(t)$ (see the next Section for
more details).

\section{Numerics vs Analytics}

In the following we are focusing on a class of models
defined by the currents
\be 
j(u) = {u\over (1+u^2)^\alpha} ,
\label{j}
\ee
which correspond to the potentials
\be
V(u) = -{1\over 2(\alpha -1)} {1\over (1+u^2)^{\alpha -1}} .
\label{V}
\ee
They were introduced in a previous paper on this journal~\cite{EPJB}
and were called $\alpha-$models.

It is straightforward to check that $V(u)$ has a minimum
in $u=0$ and goes to zero for large $u$, as 
$V(u)\sim -1/|u|^{2(\alpha-1)}$.
The curves $\lambda(A)$ can be found numerically, but the sign of 
$d\lambda/dA$ can be easily deduced from the behavior of $V(u)$
at small and large $u$. At small $u$, $V(u)\approx V(0) +
\fra{1}{2}u^2 -\fra{\alpha}{4}u^4$, so that the quartic term is negative.
At large $u$, $V(u)$ increases and goes to a constant value. 
Both these features are signatures~\cite{note_lambda} for a
positive $d\lambda/dA$, i.e. for a perpetual coarsening.

\subsection{Old results}

In Ref.~\cite{EPJB} we studied the coarsening exponent $n$
($\lambda(t)\sim t^n$) for the $\alpha-$models.
The non-conserved version, Eqs.~(\ref{gGL},\ref{j}), allowed
for an analytical treatment which followed an approach due to
Langer~\cite{Langer}: it consists in evaluating the most unstable
eigenvalue of the linear operator describing perturbations
of the periodic stationary solution. This method gave the results
\be
n = \left\{
\begin{array}{ccc}
\fra{1}{2} & & \alpha < 2  \cr
{\alpha\over 3\alpha -2} & & \alpha > 2
\end{array}
~~~\mbox{non-conserved~~(Ref.~\protect\cite{EPJB})}
\right.
\label{n_nc}
\ee

The conserved version, Eqs.~(\ref{gCH},\ref{j}), of $\alpha-$models
did not allow for an equally rigorous approach.
Therefore, we integreted numerically~\cite{EPJB}
the growth equations $\partial_t z =-\partial_x {\cal J}$:
it appeared that coarsening exponents agreed fairly well
with the relations
\be
n = \left\{
\begin{array}{ccc}
\fra{1}{4} & & \alpha < 2  \cr
{\alpha\over 5\alpha -2} & & \alpha > 2
\end{array}
~~~\mbox{conserved~~(Ref.~\protect\cite{EPJB})}
\right.
\label{n_c}
\ee

These results have been interpreted by doing the following ansatz:
``{\em as for the coarsening exponent, passing from the non-conserved 
to the conserved models is equivalent to replace $(-\partial_{xx})$
in Eq.~(\ref{gCH}) with $1/\lambda^2$.}"
This recipe allows to get the conserved coarsening exponents
(\ref{n_c}) from the non-conserved ones (\ref{n_nc}) in a
straightforward manner. The previous picture appeared to be reasonably
correct until a more general analytical approach~\cite{PRE} 
has been recently developed.

\subsection{New results}
\label{sec_IJ}

This new theory relies on the observation that
the coarsening law $\lambda(t)$ can be extracted from the
so-called phase diffusion coefficient, which describes the dynamics
of the local phase, when the periodic stationary solution is
perturbed. This approach is applicable to large classes of models,
both conserved and non-conserved. As for the $\alpha-$models,
$\lambda(t)$ is deduced from the relations~\cite{PRE}
\bea
{j(A)\over I\,\lambda'(A)} \sim {1\over t} &&
\mbox{conserved models}\label{D_gCH} \\
{j(A)\over J\,\lambda'(A)} \sim {1\over t} &&
\mbox{non-conserved models}\label{D_gGL}
\eea
In Eqs.~(\ref{D_gCH},\ref{D_gGL}), all the quantities refer
to the periodic steady states, $u(x+\lambda)=u(x)$, satisfying the
equation $u_{xx}+j(u)=0$: $A$ is the amplitude (the maximal positive
value of $u(x)$); $\lambda$ is the oscillation period (i.e. the
wavelength) and $\lambda'(A)$ is its derivative with respect to $A$; 
$J$ is the action variable, defined by $J=\oint dx \, u_x^2$; 
finally, $I=\oint dx \, u^2$.

For large $A$, we can split the motion of the fictitious particle
in the potential $V(u)$ in a region close to the origin,
$|u|<A_0$, and in the complementary regions $A_0<|u|<A$.
$A_0$ is choosen so that in the regions $|u|>A_0$, $j(u)$ and
$V(u)$ can be approximated by their asymptotic expressions
$j(u)\simeq 1/|u|^{2\alpha-1}$ and $V(u)\simeq
-\fra{1}{2(\alpha-1)}|u|^{-2(\alpha-1)}$.

Since $V(A)$ goes to a constant for diverging $A$, the motion
of the particle in the small $u$ region ($|u|<A_0$) does 
{\em not} depend on $A$.
Therefore, every integral quantity ($\lambda,I,J$) is the sum of a
constant term, coming from the integration in the small $u$ region, and an 
asymptotic $A-$dependent term, coming from the integration in the
large $u$ region. In all cases (with one exception, see below) the asymptotic
contribution diverges with $A$ and therefore dominates. Such contribution can
be evaluated by dimensional
arguments and, more rigorously, 
using the law of mechanical similarity~\cite{Landau}.
For example, $\lambda(A)$ can be simply deduced equating the
`acceleration' $A/\lambda^2$ to the `force' $j(A)\sim 1/A^{2\alpha-1}$,
therefore getting $\lambda\sim A^\alpha$.
Similarly, we get $I\sim\lambda A^2$. 

As for $J$, the asymptotic
contribution amounts to $\lambda(A/\lambda)^2\sim A^{2-\alpha}$, 
which diverges for $\alpha<2$ only. 
For $\alpha>2$ (this is the exception mentioned
above) the asymptotic contribution of $J$ vanishes, 
indicating that the
small $u$ region, giving a constant contribution, dominates.
In conclusion, $J\sim A^{2-\alpha}$ for $\alpha<2$ and $J\sim 1$ for 
$\alpha\ge 2$. 

If we replace the previous relations in Eqs.~(\ref{D_gCH},\ref{D_gGL}),
we get Eq.~(\ref{n_nc}) for the non-conserved models, but
we get a constant coarsening exponent, $n=\fra{1}{4}$, 
for the conserved models, in sharp contrast with Eq.~(\ref{n_c}). 
In conclusion, the Langer-type approach and the recent theory 
based on the phase diffusion coefficient give the same results
(\ref{n_nc}) for the non-conserved $\alpha-$models.
For the conserved models, where the Langer-type approach is
not applicable, the recent theory seems not to agree with the numerical
results (\ref{n_c}) found in Ref.~\cite{EPJB} via direct numerical
integration. This disagreement also calls for reconsidering
the ansatz $(-\partial_{xx}) ~\to ~ 1/\lambda^2$.
Next Section is devoted to understand the origins of this
discrepancy (therefore, we will limit to $\alpha>2$).

\section{The origin of the effective exponents}

First of all, let us determine numerically $\lambda(t)$ from Eq.~(\ref{D_gCH})
at all times. Results for $\alpha=3$ are shown as circles in Fig.~1 (main).
The direct integration of the growth equation was performed 
in Ref.~\cite{EPJB} up to
$\lambda\approx 10^2$. A numerical fit in this region (dashed line) gives
an effective esponent $n=0.231$, which is in perfect agreement with
such simulations and with the relation $n=\alpha/(5\alpha-2)$.
Conversely, a fit in the asymptotic region (full line) gives
$n=0.25$, as expected.
The numerical results found in \cite{EPJB} are therefore interpreted
as finite-time exponents: formulas (\ref{n_c}) are applicable up to
$\lambda\approx 10^2$, but the asymptotic exponent for the
conserved model is $\fra{1}{4}$.

\begin{figure}
\includegraphics[width=8cm,clip]{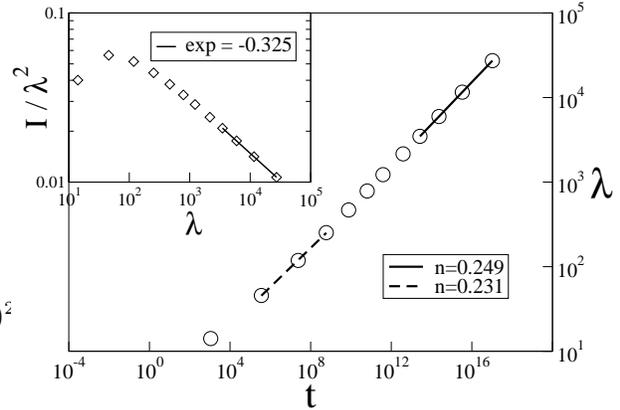}
\caption{The coarsening law $\lambda(t)$ for the conserved model
$\alpha=3$, as derived by Eq.~(\protect\ref{D_gCH}) after numerical
determination of the steady states $u(x)$. Full and dashed lines
refer to the asymptotic fit and to the fit in the region 
$\lambda\approx 10^2$, respectively. Inset: The quantity $I/\lambda^2$
as a function of $\lambda$, for the same conserved model.}
\end{figure}

Let us now discuss the origin of such finite-time exponents.
As discussed in the previous Section, formulas (\ref{n_c})
are correct insofar as the ansatz $(-\partial_{xx}) ~\to ~ 1/\lambda^2$
is correct. Its validity corresponds to say that
\be
\left({1\over t}\right)_{\mbox{conserved}} \sim {1\over\lambda^2} 
\left({1\over t}\right)_{\mbox{non-conserved}} .
\ee
If we focus on our {\it exact} relations (\ref{D_gCH},\ref{D_gGL}), 
this relation would imply 
\be
{j(A)\over I(\partial_A\lambda)} \sim
{1\over\lambda^2} {j(A)\over J(\partial_A\lambda)} ,
\ee
i.e. $I/\lambda^2\sim J$.
Since $J$ is constant for $\alpha>2$, the formula 
$n=\alpha/(5\alpha-2)$ would be correct
if $(I/\lambda^2)$ would be constant as well.
In the inset of Fig.~1 we plot the numerical results for
$I/\lambda^2$ {\it vs} $\lambda$: 
for very large $\lambda$, $I/\lambda^2\sim\lambda^{-1/3}$ decreases,
but for $\lambda\approx 10^2$ it has a {\it maximum}.
In other words, in the region of wavelengths which can be 
reasonably investigated with the direct integration of the
growth equation, the quantity $I/\lambda^2$ is approximately
constant, which implies $n\approx\alpha/(5\alpha-2)$.

\begin{figure}
\includegraphics[width=8cm,clip]{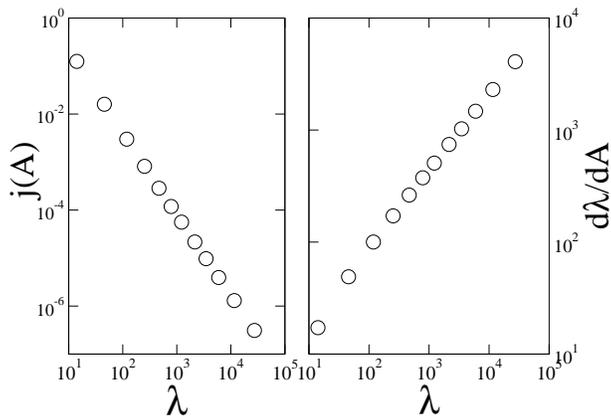}
\caption{The scaling of the maximal force $j(A)$ (left) and 
of the quantity $\lambda'(A)$ (right), as functions of $\lambda$.}
\end{figure}

In order to support the idea that the origin of the effective
exponents is indeed the maximum in $I/\lambda^2$, in Fig.~2
we also plot the two other relevant quantities appearing in
Eq.~(\ref{D_gCH}), $j(A)$ and $d\lambda/dA$: none of them
has any special behaviour for small $\lambda$.

\section{Conclusions}

In this short note we have reconsidered a class of
conserved (\ref{gCH}) and non-conserved (\ref{gGL})
growth models, in the light of recent theoretical results~\cite{PRE}.
These models, defined by the current (\ref{j}) and termed
$\alpha-$models, all display perpetual coarsening,
$\lambda(t) \sim t^n$.

For the non-conserved models, $n$ depends on $\alpha$ according
to formula (\ref{n_nc}): therefore, the theory based on the
phase diffusion coefficient~\cite{PRE} confirm previous 
results~\cite{EPJB} based on a Langer-type approach~\cite{Langer}.
For the conserved models, recent theoretical results~\cite{PRE} give a constant
coarsening exponent, $n=\fra{1}{4}$, at odd with our previous
numerical results, Eq.~(\ref{n_c}).
We have explained the effective exponent $n=\alpha/(5\alpha -2)$ as
a finite-time exponent, whose appearance is due to the fact that
the quantity $I/\lambda^2$, instead of decreasing as a power law,
is approximately constant for not too large $\lambda$ (see
Fig.~1, inset). This constant behaviour is equivalent
to assume that the operator $(-\partial_{xx})$ in Eq.~(\ref{gCH})
can be effectively replaced by $1/\lambda^2$, when $n$ is evaluated.

It is worthnoting that the result $n=\fra{1}{4}$ for the conserved models 
was firstly found by Golubovi\'c~\cite{Golubovic} using some
dimensional arguments (see Section~7 of Ref.~\cite{EPJB} for
more details). The main problem with this approach is that it also
gives a constant coarsening exponent $n=\fra{1}{2}$ for the
non-conserved models, which is wrong.
The reason of this failure is clear from Section~\ref{sec_IJ},
because dimensional analysis can be valid only if the
$A-$dependent  asymptotic contribution to the integral
quantities $\lambda,I,J$ outnumbers the contribution from the
small $u$ region. This is not the case for the
non-conserved models and $\alpha > 2$: exactly the class
of equations where dimensional analysis fails.

Similar arguments can be used to understand the failure
of the ansatz $(-\partial_{xx} \to 1/\lambda^2)$, which has clearly
to do with dimensional analysis. The weak point is that this ansatz 
is applied along with the correct results for the non-conserved models,
where dimensional analysis works for $\alpha <2$ only:
therefore, the ansatz gives the final correct result (for
the conserved models) for $\alpha <2$ only.

It is also worth stressing that a more rigorous application of 
dimensional arguments~\cite{Krug} to the conserved models 
gives an inequality, $n\le\fra{1}{4}$,
which does not allow to discriminate between  $n=\fra{1}{4}$ and
$n=\alpha/(5\alpha -2)$.

We conclude with some remarks on the possibility to
access numerically the asymptotic scaling region reported in
Fig.~1 for the conserved model with $\alpha=3$.
We are going to argue that a direct integration 
of Eq. (\ref{eq_z}) 
would require astronomically long CPU times.

Since the correct scaling sets in
for $\lambda > 2000$ (see Fig.~1), it would be necessary to consider
a chain of length $L \sim 10^4$ and to integrate for
times $t \sim 10^{15}$.
In our numerical results published in Ref.~\cite{EPJB}, we integrated 
Eq.~(\ref{eq_z}) using a time-splitting pseudo-spectral code, using
a spatial resolution $\Delta x = 0.25$, a time step $\Delta t = 0.05$
and a chain of length $L=1024$.
Employing an ``Opteron AMD64 Dual Core" machine
with a 2 GHz clock, we are {\em currently} 
able to reach $\lambda\approx 10^2$ 
and $t\approx 10^8$ with 20 hours of CPU time.
This means that in order to simulate
a chain of length $L \sim 10^4$ for a time $t \sim 10^{15}$, we would
require (on the same machine) a CPU time of
$10\times 10^7\times 20$ hours $\approx 2 \times 10^5$ years.
The simulation time can be reduced to some extent by lowering the
precision of the integration. In particular, 
we have verified that results of quality comparable with those reported in~\cite{EPJB} 
can still be obtained by increasing the time step up to four times, while
the integration scheme becomes rapidly unstable by considering
a coarser space grid. As a matter of fact, we cannot expect to lower the
CPU time more than a factor $\sim 10$, which renders still unfeasible 
the observation of the asymptotic exponents
(by the way, even a lowering up to a factor $10^5$ would 
make it unfeasible).


\begin{thebibliography}{}
\bibitem{review}
P. Politi, G. Grenet, A. Marty, A. Ponchet, J. Villain,
Phys. Rep. {\bf 324}, 271 (2000).
\bibitem{libro_JV}
A. Pimpinelli and J. Villain, {\it Physics of Crystal Growth}
(Cambridge University Press, Cambridge, 1998).
\bibitem{proceeding}
P. Politi, J. Villain, 
in {\it Surface diffusion: atomistic and
collective processes}, ed. M.C. Tringides
(Plenum Press, New York, 1997), page 177.
\bibitem{Evans}
J.W. Evans, P.A. Thiel, M.C. Bartelt,
Surf. Sci. Rep. {\bf 61}, 1 (2006).
\bibitem{libro_MK}
T. Michely, and J. Krug,
{\em Islands, Mounds and Atoms} (Springer, Berlin, 2004).
\bibitem{PRE}
P. Politi, C. Misbah, Phys. Rev. E {\bf 73}, 036133 (2006).
\bibitem{PRL}
P. Politi, C. Misbah, Phys. Rev. Lett. {\bf 92}, 090601 (2004).
\bibitem{EPJB}
A. Torcini, P. Politi, Eur. Phys. J. B {\bf 25}, 519 (2002).
See also: P. Politi and A. Torcini, 
J. Phys. A: Math. Gen. {\bf 33}, L77 (2000).
\bibitem{Langer}
J. S. Langer, Ann. Phys. {\bf 65}, 53 (1971).
\bibitem{note_lambda}
See Ref.~\protect\cite{PRE}, Appendix A.
\bibitem{Landau}
L.D. Landau and E.M. Lifshitz, {\it Mechanics}
(Pergamon Press, Oxford, 1976), Section 28.
\bibitem{Golubovic}
L. Golubovi\'c, Phys. Rev. Lett. {\bf 78}, 90 (1997).
\bibitem{Krug}
M. Rost and J. Krug, Phys. Rev. E {\bf 55}, 3952 (1997).
\end{thebibliography}
\end{document}